# Urban-rural gap and poverty traps in China: A prefecture level analysis

Jian-Xin Wu[a], Ling-Yun He[a, b, 1]

[a] School of Economics, Jinan University, Guangzhou 510632, China

[b] College of Economics and Management, Nanjing University of Information Science and Technology, Nanjing 210044, China

**Abstract**: Urban-rural gap and regional inequality are long standing problems in China and result in considerable number of studies. This paper examines the dynamic behaviors of incomes for both urban and rural areas with a prefectural data set. The analysis is conducted by using a distribution dynamics approach, which have advantages in examination on persistence, polarization and convergence clubs. The results show that persistence and immobility are the dominant characteristics in the income distribution dynamics. The prefectural urban and rural areas converge into their own steady states differentiated in income levels. This pattern of urban-rural gap also exists in three regional groups, namely the eastern, central and western regions. Examination on the dynamics of the poorest areas shows that geographical poverty traps exist in both urban and rural prefectural areas. Our results indicate that more policy interventions are required to narrow down the urban-rural gap and to eliminate the poverty traps in China.

---

[1] Corresponding author. Dr. Ling-Yun He is a full professor in applied economics in the School of Economics at Jinan University, and an affiliate professor in the College of Economics and Management at Nanjing University of Information Science and Technology, email: lyhe@amss.ac.cn. Dr. Jian-Xin Wu is an associate Professor in applied economics in the School of Economics at Jinan University, twujianxin@jnu.edu.cn. This work was supported by the China National Social Science Foundation (No. 15ZDA054), Humanities and Social Sciences of Ministry of Education Planning Fund (16YJA790050) and the National Natural Science Foundation of China (Nos. 71473105, 71273261 and 71573258).

# Urban-rural gap and poverty traps in China: A prefecture level analysis

**Abstract**: Urban-rural gap and regional inequality are long standing problems in China and result in considerable number of studies. This paper examines the dynamic behaviors of incomes for both the prefectural urban and rural areas. The analysis is conducted by using a distribution dynamics approach, which have advantages in examination on persistence, polarization and convergence clubs. The results show that persistence and immobility are the dominant characteristics in the income distribution dynamics. The prefectural urban and rural areas converge into their own steady states differentiated in income levels. This pattern of urban-rural gap also exists in three regional groups, namely the eastern, central and western regions. Examination on the dynamics of the poorest areas shows that geographical poverty traps exist in both prefectural urban and rural areas. Although the Chinese government has launched many programs to reduce income inequality, our results indicate that more policy interventions are required to narrow down the urban-rural gap and to eliminate the poverty traps in China.

## 1. Introduction

China's rapid economic growth since the reform is accompanied with significant increase in income inequality. The high and rising income inequality has result in many adverse effects on the social and economic development. It dampens domestic consumption, contributes to the trade balance, and undermines social cohesion and political stability (Wang et al., 2014). Moreover, high income inequality weakens the poverty reduction effect of economic growth. This is easy to understand, the same growth rate can reduce more poverty in a more equal society than in an unequal society. As indicated by Wan and Sebastian (2011), more than 100 million Chinese surviving on no more than 2.25 USD a day in 2008. Poverty reduction is still a long lasting objective for the Chinese government. Therefore, the issues of income disparity in China have attracted many attentions from the policy makers and researchers.

Income inequality has been ranked among the top socioeconomic issues for many years. Chinese government has launched many programs to reduce regional income disparity, such as the three regional development programs, namely the Western Development Program in 2000, the Rise of Central China Program, and the Revitalization of the Northeast Program in 2004. The Socialist New Countryside Development Program and abolishment of agriculture taxes were launched in 2000 and 2005 respectively, which aimed at tackling the urban-rural gaps. More recent efforts include the goal of "Building a Harmonious Society" in the 11$^{th}$ Five Year Plan, and the expansion of social protection to the rural population. However, whether these policy interventions are effective in terms of poverty reduction still remains an open question.

In current literature, there are some studies on Chinese income disparity. These studies usually focus on three dimensions, namely interhousehold inequality, regional disparity, and urban-rural gap. Due to data availability, the studies on interhousehold inequality are based on inconsecutive household survey data collected by the Chinese academy of social sciences and other unofficial organizations. These data sets only cover partial provinces and have relatively small sample sizes. Most of the studies on regional disparity and urban-rural divide use provincial panel data. Few of them use more disaggregated data such as prefecture or county level data. Furthermore, index approach and parametric approach are most popular in these studies. However, these approaches only provide the income behavior of an average or representative economy (Quah, 1996a), instead of critical information for intradistribution dynamics, such as catch-up effect, polarization, convergence clubs, and poverty traps.

This paper, therefore, aims at investigating the distribution dynamics of Chinese prefectural urban and rural incomes for the most recent period 1999-2013. This paper differs from previous studies in several ways. First, we use a prefectural urban and rural income data to investigate both inter-provincial and intra-provincial disparities. Moreover, the sample size of this data set is more than ten times larger than that of a provincial one. Second, we use a distribution dynamics approach, which has several advantages over the traditional index approach and parametric approach, to examine the evolution of regional incomes. This distribution dynamics approach can provide information for the law of the motion of the entire shape of the income distribution. In particular, it can provide insights on persistence, convergence clubs, and poverty traps. Third, this paper examines the income distribution dynamics of both urban and rural areas. This may provide us a more comprehensive understanding of the evolution trend of Chinese spatial income distribution from two perspectives, namely regional inequality, and urban-rural gap.

The remainder of the paper is organized as follows. Section 2 provides a brief review of the related literature. Section 3 introduces the methodology used in this paper. Section 4 describes the data. Section 5 presents the main empirical results. Section 6 concludes.

## 2. Related literature

Our research is closely related to two categories of literature. The first category is the studies on the dynamic evolution of inequality in China. The second focuses on the poverty traps in China.

Chinese regions differ greatly in geographical location, factor endowment, and governmental policies. As most of Chinese minorities live in some poor inland areas, regional disparity is considered to be closely related with ethnic tensions. High interregional inequality may undermine national unity. Therefore, interregional disparity has attracted persistent research interests. These studies include Lardy (1978), Lyons (1991), Tsui (1991), Kanbur and Zhang (2005), Zhang and Zou (2012), etc. Due to data availability, most of these studies examine the interregional disparity and urban-rural gap using provincial data. Instead of regional disparity, some of the researches focus on the urban-rural gaps in China, such as Nolan (1979), Kwong (1994), Zhao (1993), Yang and Zhou (1999), Wang et al. (2016). Almost all researches indicate that China has very large urban-rural income gaps, and that these gaps are continually expanding. The urban-rural gap in China has a significant regional dimension and exists in all provinces in China. Moreover, urban-rural gap contribute more to the general disparity in China than regional inequality.

In recent years, some household survey data became available, such as the data set collected by Chinese Academy of Social Sciences (CASS) in 1986. And the data set collected by the Chinese Household Income Project (CHIP) for 1988, 1995, 2002 and 2007. The researches based on the household data includes Wan and Zhou (2005), Demurger et al. (2006), Sicular et al. (2007), Ravallion and Chen (2007), Li et al. (2013), etc.. These household data sets provide possibility to examine the interhousehold income disparities. As interhousehold disparity encompasses regional and urban-rural inequality, most of these studies decompose the interhousehold disparity into regional and urban-rural inequality. These findings further confirm that urban-rural gap is the dominant component in Chinese general inequality. However, these household survey data suffer from small sample sizes and time inconsecutiveness.

The poverty trap is the other focus in the current literature of China study. The concept of poverty trap means that poverty begets poverty, and that current poverty per se is a direct cause of poverty in the future. This self-reinforcing mechanism causes poverty to persist (Azariadis and Stachurski, 2005). Jalan and Ravallion (2002) made the pioneering effort to examine poverty trap in China, and found robust evidence of geographical poverty trap in rural China. Using household survey data in 2002, Knight et al. (2009, 2010) examined the effect of education in the formation of poverty trap in rural China, verified the existence of education trap and poverty trap, and described the mechanism how these two traps reinforce each other. Cao et al. (2009) investigated the roles played by the urban-biased policies in the formation of poverty trap in rural China. Using China Health and Nutrition Survey (CHNS) household data and a multilevel econometric model, Fang and Zou (2014) explored how neighborhood effects are interrelated with the chronic poverty, and found that group variables have significant impact on the individual incomes. However, all these studies focus on the poverty trap in rural areas. As indicated by Cheng et al. (2002) and Wan (2007), with the fast rising inequality in China, urban poverty also emerges as well as rural poverty. Both intra-urban and intra-rural inequalities have been increasing significantly in recent years (Wang et al., 2014). Therefore, studies on poverty in both urban and rural areas can shed more light on the evolution dynamics of income disparity in the transitional China.

## 3. Research method

Traditional studies on income inequality often use Theil index or Gini coefficient to analyse the evolution of income inequality, or apply various inequality decomposition techniques to examine the components in the general inequality. These approaches provide a general measure of total inequality or its components, and the income behaviour of an average or representative economy. However, it sheds little light on the dynamic evolution of each part, or even each sample, such as the polarization, poverty trap and convergence clubs. Alternatively, this paper adopts a distribution dynamics approach developed by Danny Quah in a series of papers (Quah, 1993, 1996a, 1996b, 1997). In comparison to the traditional approaches, the distribution dynamics approach has two advantages. First, it can provide insight on the law of motion in an entire shape of the income distribution, particularly in the formation of convergence clubs, poverty trap and polarization. Second, this approach is completely data-driven, and imposes no assumptions on the model. This may be helpful to avoid the estimation bias due to model assumptions in the traditional parametric approaches.

In empirical studies, two distribution dynamics methods, namely the discrete and continuous approaches, are widely used. Most of the early studies use the discrete approach due to simplicity in calculation. However, the discrete approach suffers from the problem of arbitrarily discretising the variable of interest into different state spaces (mostly 5 or 7 states). The discretisation process may change the probabilistic properties of the variable in question. Moreover, some studies, such as Quah (1997), Bulli (2001), and Johnson (2005), argue that the estimation results of the discrete approach are sensitive to this discretisation process. However, the continuous distribution dynamics approach adopts a nonparametric stochastic kernel approach to estimate the distribution with infinite state spaces. Thus it avoids the problem of discretization, and can be considered as an upgraded version of the discrete approach. This paper uses the continuous version of the distribution dynamics approach to examine the dynamic behaviour of income distributions across Chinese prefectural urban and rural areas.

Suppose that distribution of prefectural income $x$ at time $t$ can be denoted by the density function $f_t(x)$. The evolution of the distribution is time-invariant and first-order, that is, the current distribution $f_t(x)$ at time $t$ will evolve into the future distribution $f_{t+\tau}(y)$ at time $t+\tau$, where $\tau>0$. Then the relationship between the current distribution $f_t(x)$ and the future distribution $f_{t+\tau}(y)$ can be described as:

$$f_{t+\tau}(y) = \int_0^\infty g_\tau(y|x) f_t(x) dx \qquad (1)$$

where $g_\tau(y|x)$ is the conditional probability density function that maps the transition process of the distribution of income from time $t$ to time $t+\tau$. Keeping the conditional probability density function $g_\tau(y|x)$ steady, the distribution of prefectural income will evolve into a long-run equilibrium state named as the ergodic distribution $f_\infty(y)$. With this definition, the ergodic distribution $f_\infty(y)$ can be estimated by:

$$f_\infty(y) = \int_0^\infty g_\tau(y|x) f_\infty(x) dx \tag{2}$$

This paper uses a kernel density approach to estimate the conditional probability density function $g_\tau(y|x)$ and the long-run equilibrium distribution $f_\infty(y)$. First, we define the joint natural kernel estimator of $f_{t,t+\tau}(y,x)$ and the marginal kernel estimator of $f_t(x)$ as:

$$f_{t,t+\tau}(y,x) = \frac{1}{nh_x h_y} \sum_{i=1}^n K(\frac{x-x_i}{h_x}, \frac{y-y_i}{h_y}) \tag{3}$$

$$f_t(x) = \frac{1}{nh_x} \sum_{i=1}^n K(\frac{x-x_i}{h_x}) \tag{4}$$

where $x_i$ is the income of a specific prefecture at time $t$, and $y_i$ is the income value of that prefecture at $t+\tau$. We use $K(\cdot)$ to denote kernel function. $h_x$ and $h_y$ stand for the bandwidths of $x$ and $y$ respectively. Following Quah (1997) and Johnson (2005), the bandwidths are estimated using the method described in Silverman (1986).

With above definitions, we can further estimate the conditional probability density with $f_{t,t+\tau}(y,x)$ and $f_t(x)$ through:

$$g_\tau(y|x) = \frac{f_{t,t+\tau}(y,x)}{f_t(x)} \tag{5}$$

Three-dimensional plot and contour map are the most popular tools to show estimation results of transition probability in the continuous approach. The three-dimension plot shows the entire shape of the transition probability distribution, while contour plot illustrates the deviation from the diagonal. To gain more insights into the transition probability mass, we also estimate the net transition probability (NTP) values at each point to show the general mobility. Following Cheong and Wu (2014), the net transition probability index $p(x)$ can be estimated by:

$$p(x) = \int_x^\infty g_\tau(z|x)dz - \int_0^x g_\tau(z|x)dz \quad (7)$$

Intuitively, a positive net transition probability value at a point indicates an increasing trend in income, while a negative net transition probability value at a point implies a decreasing trend. In particular, if a net transition probability curve has a negative slope and only one intersection with the horizontal axis, this implies that the poor grows faster and the rich slower. This indicates the existence of simple convergence.

## 4. Data description

This paper uses prefecture level income data set. To the best of our knowledge, prefecture level data is seldom employed to examine the income inequality in China. This data set provides net income in rural areas and disposable income in urban areas for 390 Chinese prefectures (including cities at prefecture level, and cities, counties and districts directly under the jurisdiction of provincial government). For simplicity, we use prefectures to denote these prefecture level administration units. In comparison to the provincial data, the prefecture level data are more disaggregated. Moreover, it enlarges the sample size ten times more than that of the provincial data set. The use of the prefecture level panel data can provide more information for the dynamic behavior of the spatial income distribution, and make the empirical results more rigorous and robust.

Our data are taken from the China Statistical Yearbook for Regional Economy (various years). The nominal income is deflated to real income by provincial CPI index. This data set covers the period 1999-2013, which is broadly cover most policies aiming at reducing income inequality in China, such as the Western Development Program in 2000, the Rise of Central China Program and the Revitalization of the Northeast Program in 2004, the Socialist New Countryside Development Program in 2000, the abolishment of agriculture taxes in 2005, and the goal of 'Building a Harmonious Society' in 2006. Therefore, in can provide an empirical verification of the policy interventions.

Following common practice in distribution dynamics approaches, we use relative income (RI) which is each prefecture's income divided by its yearly average. In this paper, we pool prefectural urban and rural income samples together for the purpose of comparison. Therefore, the relative urban income (RUI) or relative rural income (RRI) value of a specific prefecture correspond to the times of the income for this prefecture's urban or rural area relative to the sample average. For simplicity, in the following sections, the income refers to relative income.

## 5. Preliminary analysis

Figure 1 shows the kernel density distribution of rural income and urban income in the initial and final years separately. The income distributions of urban and rural areas in the final year are broadly similar to their corresponding distributions in the initial year. The peaks of urban and rural income distribution locate in the 0.5 and 1.2 times sample average respectively in the initial year. These two peaks only shift slightly to the left in our sample period. The income distributions of prefectural urban and rural areas are significantly separated. High urban-rural income gap can be observed in the distribution. Both peaks increased slightly for the period 1999-2013, suggesting some degree of convergence in the income. The distribution of rural income is more compact than that of urban income, implying that the interregional gap of urban income is higher than that of rural income. However, we can observe that the shapes of both income distributions of the urban and rural areas remain broadly stable for the sample period. This may imply persistence and immobility in relative position changes for the sample period. In general, no significant catch-up effect can be implied from the income distributions of the prefectural urban and rural areas in Figure 1.

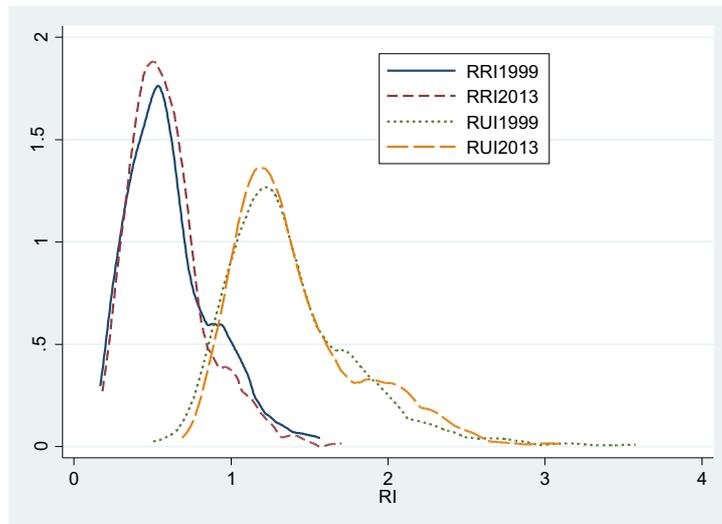

Figure 1 The kernel density distribution of RRI and RUI for the initial and final year

## 5. The main empirical results

### *5.1 The distribution dynamics of full sample*

Figure 2 shows the distribution dynamics of the pooled samples of the urban and rural income for the period 1999-2013.

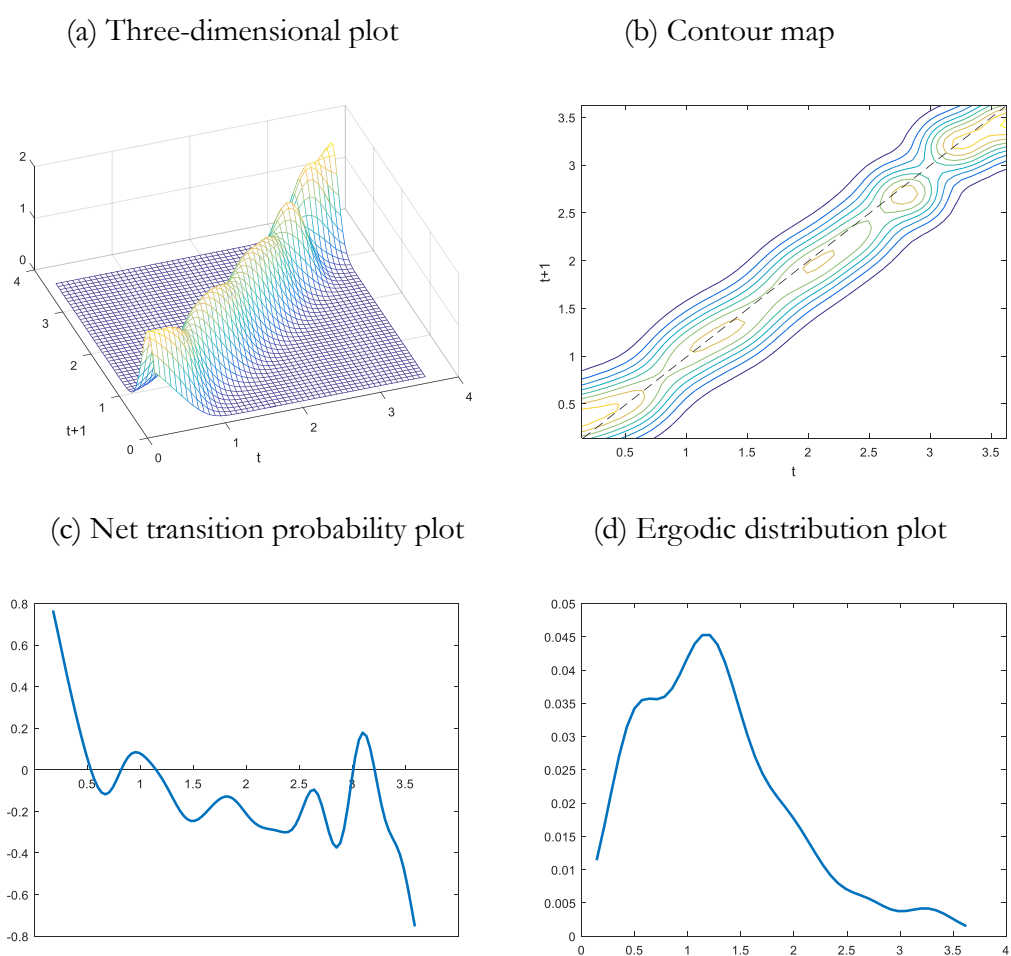

Figure 2 The distribution dynamics of pooled samples of urban and rural incomes

across Chinese prefectures for the period 1999-2013

The panels (a) and (b) of Figure 2 show the three dimensional plot and contour map of the transitional probability of the relative income. Intuitively, suppose that we chose a point 2 in axis marked *t* and slice the three-dimensional plot parallel to axis marked *t*+1, this slice indicates the probability density distribution of this area transition to other positions in the period *t*+1. Therefore, a ridge in the kernel density distribution along the 45-degree diagonal indicates strong tendency of persistence and immobility in relative position changes. The income distributions of these prefectural urban and rural areas tend to remain where they begin. On the other hand, the deviation from the diagonal shows high mobility and low persistence. In particular, the distribution of transition probability density parallels to the axis marked *t* indicates tendency of convergence. In this case, the poor regions grow faster and the rich ones slow down, and finally all regions cluster to the same income level. Finally, the existence of distinct peaks in the probability distribution implies convergence clubs differing in income levels.

Both three-dimensional plot and contour map shows that transition probability density are distributed along the diagonal. Therefore, the dominant characteristic of prefectural urban and rural income distribution is persistence and immobility. Several distinct probability density peaks can be observed along the diagonal. This implies the possible existence of convergence clubs in income distribution.

The net transition probability plot (the panel (c) of Figure 2) shows that most regions above RI value of 0.55 have negative net transition probabilities except two small regions around the average and 3.1 times average income. In addition, these two regions have relatively low positive net transition probabilities. This implies that upward tendency in these regions are not very strong, downward tendency dominant in the high income end. However, no simple trend of convergence can be implied from the net transition probability plot.

The long run distribution (the ergodic distribution plot in the panel (d) of Figure 2) is slightly multimodal. Keeping the transitional dynamics unchanged, the incomes of Chinese prefectural urban and rural areas will cluster into income groups in the future. However, we can't judge whether the ergodic distribution of the urban income is separated from that of the rural incomes. To answer this question, we need to examine the distribution dynamics of the urban and rural incomes in different sample groups.

*5.2 The distribution dynamics in the urban and rural income groups*

As indicated in previous section, the urban-rural gap in China is significantly high. The dynamic behavior of the urban and rural incomes may be different. In this sub-section, the analysis is conducted on urban and rural incomes separately.

Figure 3 shows the distribution dynamics of the incomes for the prefectural urban areas for the period 1999-2013. Similar to that in the pooled samples, the transition probability density is distributed along the 45 degree diagonal as depicted in the three-dimensional plot and contour map (the panel (a) and panel (b) in Figure 3). Several distinct density peaks exist in the high income end of the distribution. This implies persistence and immobility are the dominant characteristic in the transition dynamics of prefectural urban incomes. The poor urban areas tend to remain poor, while the rich ones tend to remain rich. In the net transition probability plot of the panel(c) in Figure 3, there are two regions have positive net transition probabilities, namely the region below the RI value of 1.25 and the region between the RI value of 3 and 3.2. However, more regions above the RI values of 1.25 have negative net transition probabilities in the higher income end.

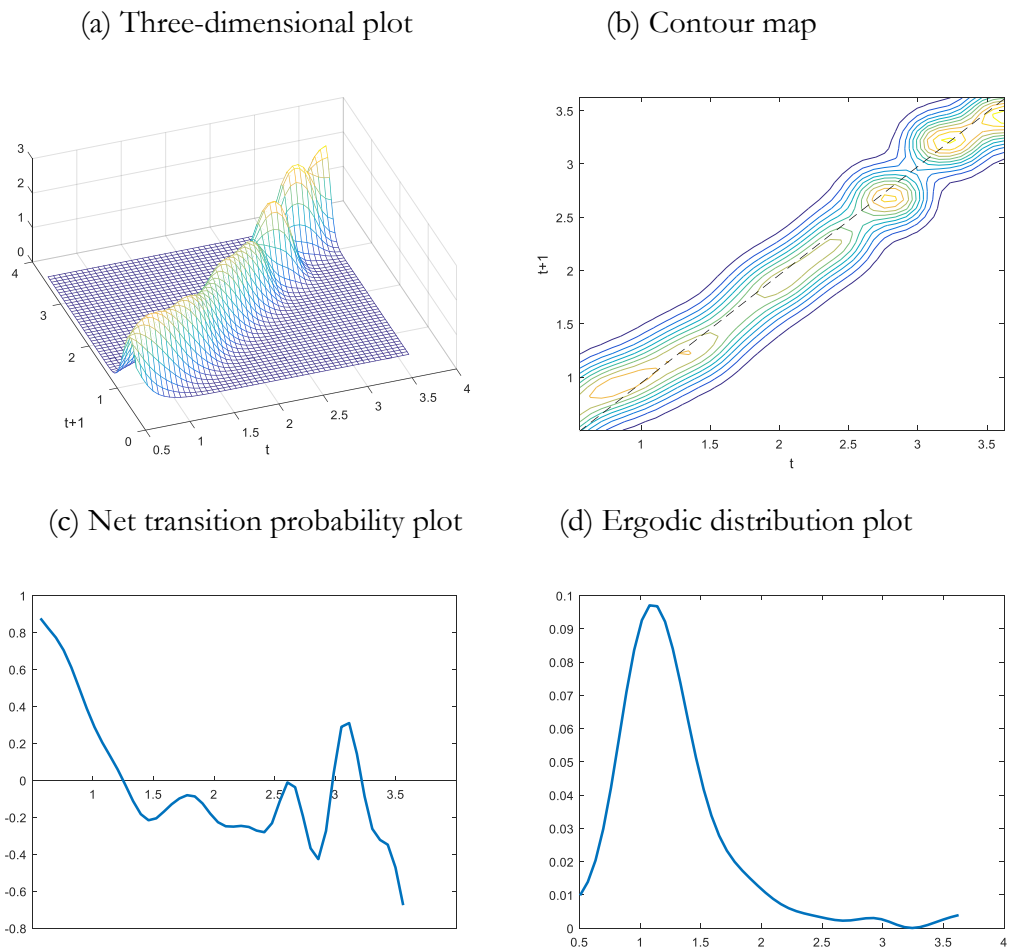

Figure 3 The distribution dynamics of the incomes for prefectural urban area for the period 1999-2013

The ergodic distribution of incomes for Chinese prefectural urban areas is significantly unimodal, only one right-skewed peak locates around the region with 1.1 times average income. Thus, the shape and position of long-run distribution is similar to that in the current distribution in Figure 1. This further indicates that persistence and immobility are the dominant characteristics in the income distribution dynamics of prefectural urban areas.

Figure 4 shows the income distribution dynamics of Chinese prefectural rural areas for the period 1999-2013. Although the transition probability density is distributed along the diagonal, the density tends to be more in the lower side than in the higher side. The net transition probability plot in the panel (c) of Figure 4 shows that regions with the RI values above 0.48 have negative net transition probabilities, while the regions with RI values below 0.48 have positive net transition probabilities. This indicates that the poorer rural areas grow faster than the richer rural areas. The catch-up effect is significant in the income distribution dynamics of Chinese prefectural rural areas. This also implies simple convergence in terms of income across the prefectural rural areas.

(a) Three-dimensional plot　　　　　　(b) Contour map

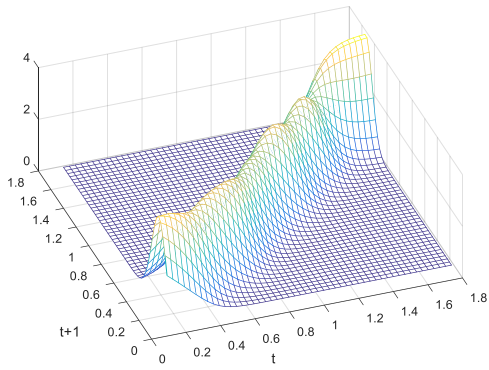 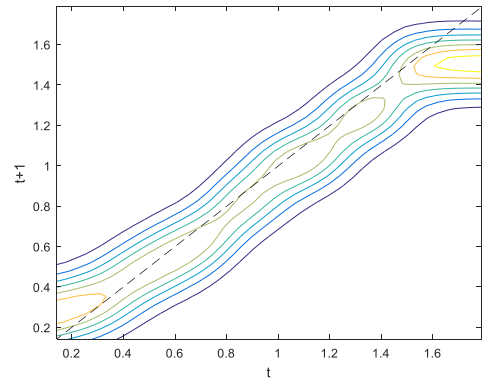

(c) Net transition probability plot    (d) Ergodic distribution plot

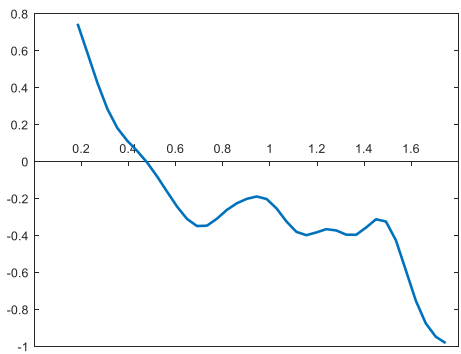 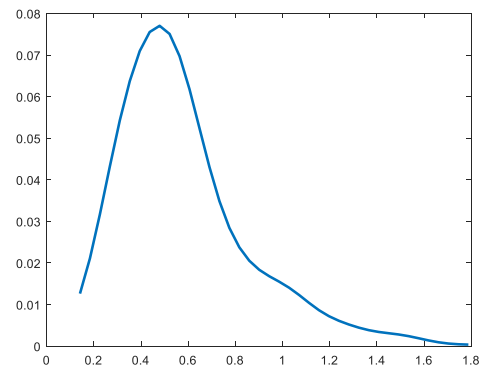

Figure 4 The distribution dynamics of the income for the prefectural rural areas for the period 1999-2013

However, the rural incomes do not converge to high income level. Instead, they converge to low level of 0.48 times average income. In comparison to the ergodic distribution of the prefectural urban areas, rural incomes and urban incomes converge to their own steady state. No significant reduction in urban-rural gap can be observed in the long run. This is somewhat discouraging. The government have conducted several intervention polices, such as the Socialist New Countryside Development Program in 2000 and abolishment of agriculture taxes in 2005, etc., to narrow down the urban-rural disparity. However, our results show that these policies have no significant effects in reducing the urban-rural gap. Therefore, more policy interventions are required to promote the economic growth in the rural areas.

***5.3 The distribution dynamics of the incomes in regional groups***

Chinese regions differ greatly in factor endowments and policies. One of the most important components of Chinese regional inequality is the east-central-west divide (Wang et al., 2014). In this section, we divide Chinese prefectures into three groups, namely the eastern, central and western groups. The income distribution dynamics of these three regions are reported as follows:

Figure 5 shows the distribution dynamics of the pooled income samples of the eastern prefectural urban and rural areas for the period 1999-2013. Strong persistence and immobility can be observed in both the three-dimensional plot and contour map (the panels (a) and (b) in Figure 5). The relative position changes among the eastern prefectural urban and rural areas are very small. No significant catchup effect or polarization can be observed in these two plots. The net transitional probability plot in the panel (c) of Figure 5 shows that most regions above the RI values of 0.85 have negative net transition probabilities except a small region around RI value of 3.1. Downward moving tendency is stronger than upward tendency in the region above the average time. The long-run distribution of the pooled urban and rural incomes of the eastern prefectures (the panel (d) in Figure 5) is almost unimodal with the peak situated around the average income.

(a) Three-dimensional plot          (b) Contour map

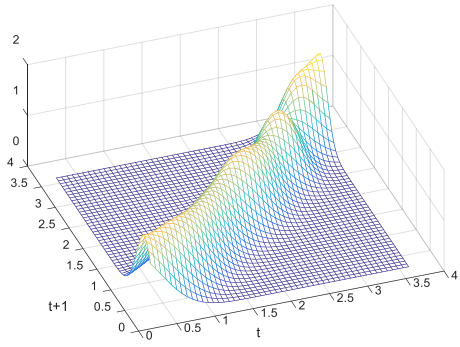 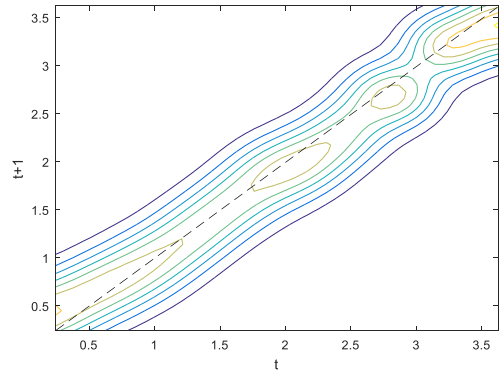

(c) Net transition probability plot     (d) Ergodic distribution plot

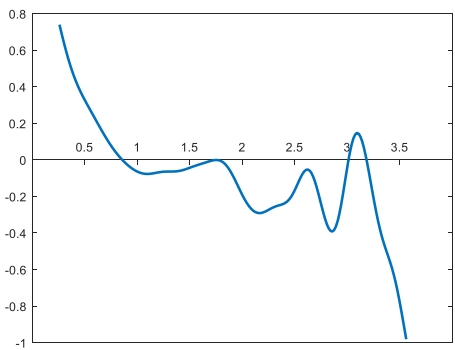 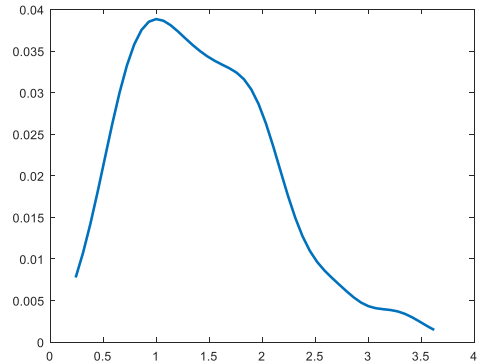

Figure 5 The distribution dynamics of the incomes for the eastern prefectural urban and rural areas for the period 1999-2013

For the purpose of comparison, we estimate the net transition probability and ergodic distribution of the eastern prefectural urban and rural incomes separately and plot them in the panels (a) and (b) in Figure 6. The net transition probability plot shows that the incomes of the eastern prefectural rural areas have strong tendency of simple convergence. Moreover, the ergodic distribution shows that the incomes of the eastern prefectural rural areas converge to the unique peak around the RI value of 0.7. On the other hand, the incomes of the eastern urban areas have much less tendency of convergence than rural income. And the ergodic distribution of the eastern urban incomes has a peak around the RI value of 1.4, which is two times that of the rural. This implies that the eastern urban and rural incomes converge to separate clubs with different income levels. Therefore, pooling the urban and rural incomes together may disguise this fact.

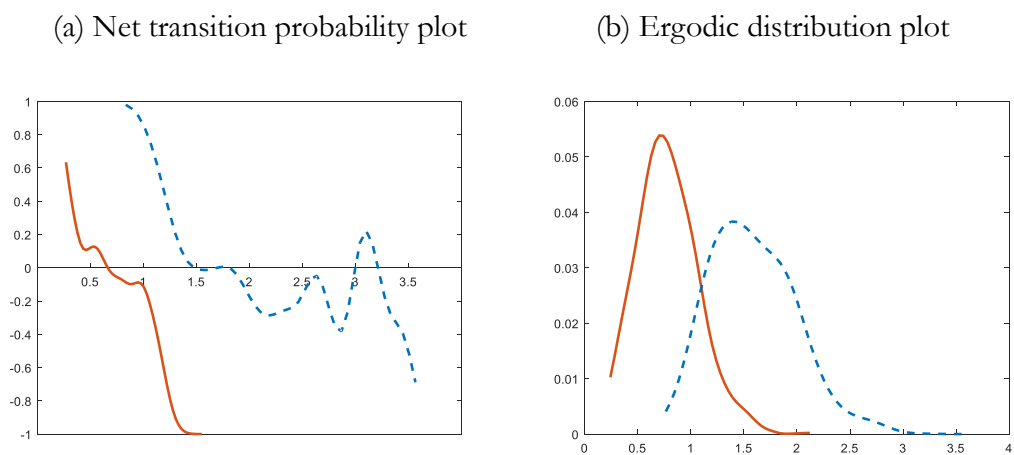

Figure 6 The urban and rural income distribution dynamics of the eastern prefectures

for the period 1999-2013

(Note: The dash line and solid line denote the urban and rural income, respectively.)

The net transition probability curve in the panel (c) of Figure 7 shows three regions with positive net transition probabilities. No simple convergence can be observed in the distribution. The ergodic distribution in the panel (d) of Figure 7 shows significant bimodality, with one peak around the RI value of 0.55, and the other around the RI value of 1.25. This indicates that keeping the distribution dynamics remain unchanged, convergence clubs exist across the pooled samples of the central prefecture urban and rural areas in the long-run.

Figure 7 plots the distribution dynamics of the pooled urban and rural incomes for central prefectures for the period 1999-2013. The probability density is distributed along the 45 degree diagonal in the panels (a) and (b) in Figure 7. This implies that persistence and immobility dominate in the intradistribution dynamics of the central prefectures. In addition, several distinct peaks can be observed in the distribution of transition probability. This implies the possible existence of convergence clubs in the long run.

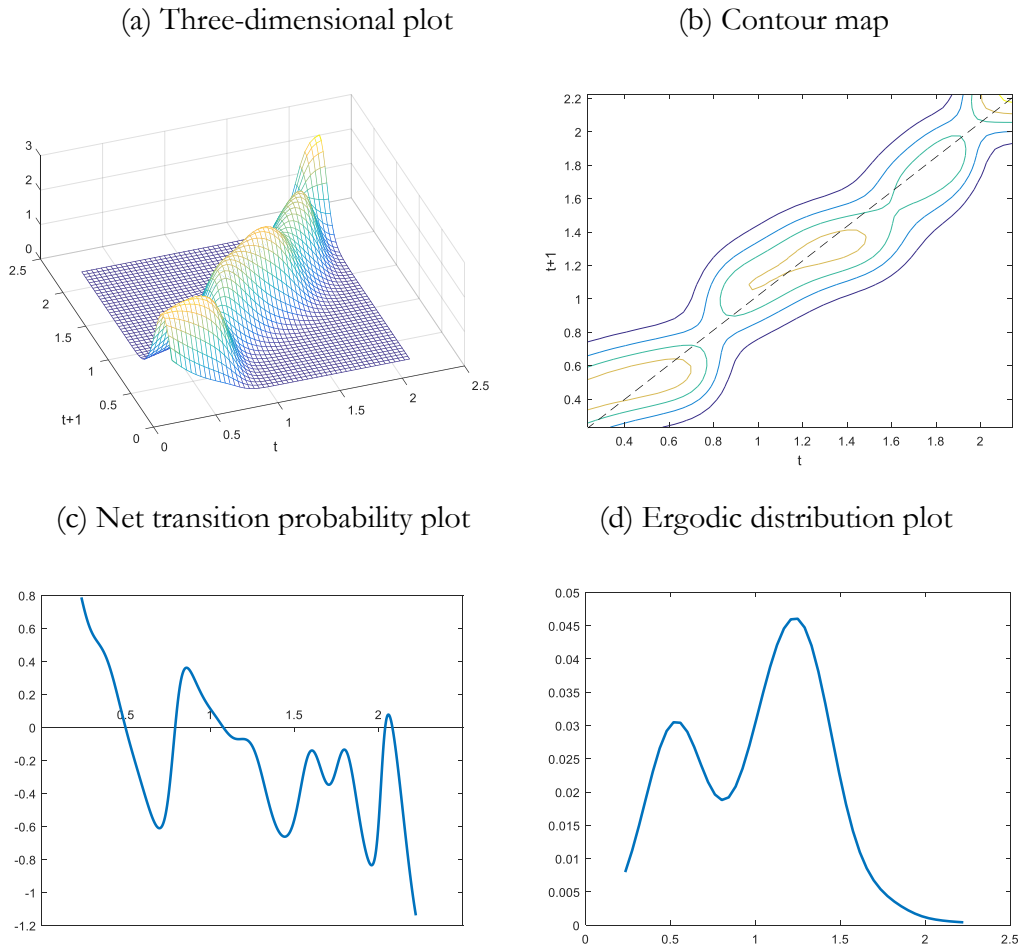

(a) Three-dimensional plot  (b) Contour map

(c) Net transition probability plot  (d) Ergodic distribution plot

Figure 7 The distribution dynamics of the incomes for the central prefectural urban and rural areas for the period 1999-2013

In comparison, Figure 8 shows the net transition probability and the ergodic plots for both urban and rural incomes estimated respectively. Similar to that in the eastern, the net transition probability plot shows that the incomes of the central prefectural rural areas have a significant tendency to converge. However, it does not converge to the average or high income. Instead, it converges to low income level of 0.4 times average.

On the other hand, the incomes of the prefectural urban areas in the central region converge to the unique peak around the average income level. This implies that the urban and rural incomes of the central prefectures converge to different income levels. Different from that in the eastern prefectures, the ergodic distributions of urban and rural incomes for the central prefectures have much less overlap. This suggests that the urban-rural gap in the central region is wider than that in the eastern region.

(a) Net transition probability plot          (b) Ergodic distribution plot

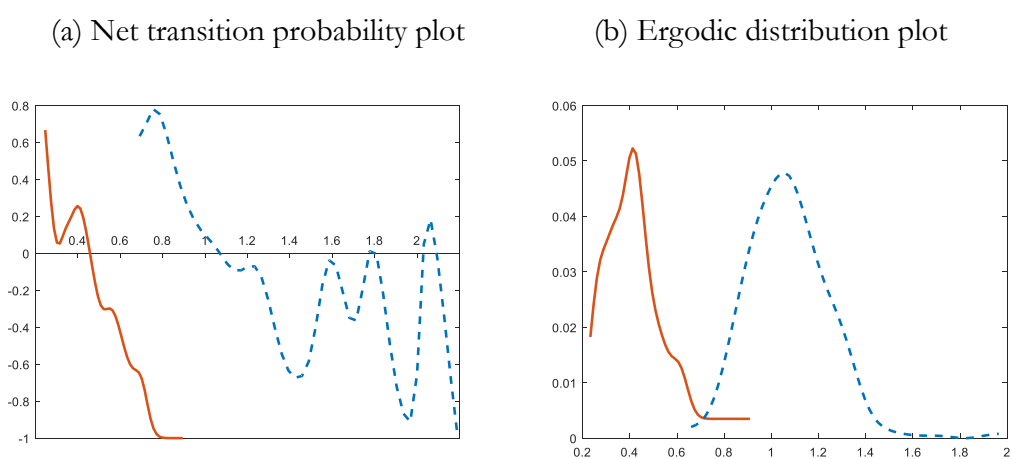

Figure 8 The urban and rural income distribution dynamics of the central prefectures for the period 1999-2013

(Note: The dash line and solid line denote the urban and rural income, respectively.)

Figure 9 plots the income distribution dynamics of the western prefecture urban and rural areas for the period 1999-2013. Both the three-dimensional plot and contour map (the panels (a) and (b) in Figure 9) show that transition probability density is distributed along the diagonal. This implies significant persistence and immobility in the intra-distribution dynamics. Three positive regions, namely below the RI value of 0.35, the region between 0.75-1.3, and the region around 2.1, can be observed in the net transitional probability plot. Thus no simple convergence can be observed from the net transition probability plot. The long-run distribution in the panel (d) of Figure 9 is significantly bimodal, with a small peak around the RI value of 0.4 and a large peak around the RI value of 1.4. This implies the existence of convergence clubs in income across the western prefectural urban and rural areas.

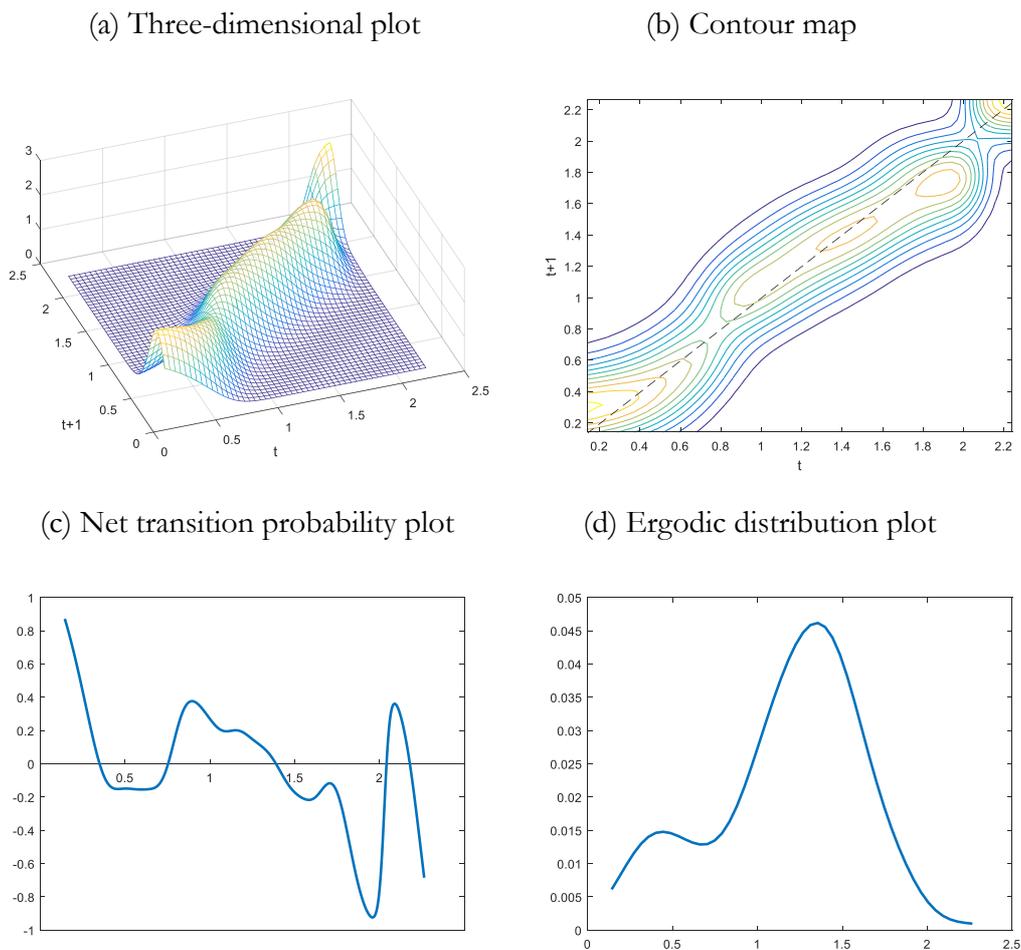

Figure 9 The income distribution dynamics of the western prefectures for the period 1999-2013

To explore whether the western prefectural urban and rural incomes converge to different income levels, we estimate net transition probability curves and the ergodic distributions of the urban and rural incomes separately and plot them in Figure 10. The net transition probability curve of the western prefectural rural incomes has a broadly negative slope and only one intersection point with horizontal axis. This indicates significant simple convergence in rural incomes of the western prefectures. However, the western urban income shows no evidence of simple convergence in the net transition probability plot. The ergodic distribution in the panel (b) of Figure 10 shows that urban and rural incomes converge to separate peaks in the long-run. The former cluster into the peak around 0.7 times average income, while the latter cluster into the peak around 1.4 times average income.

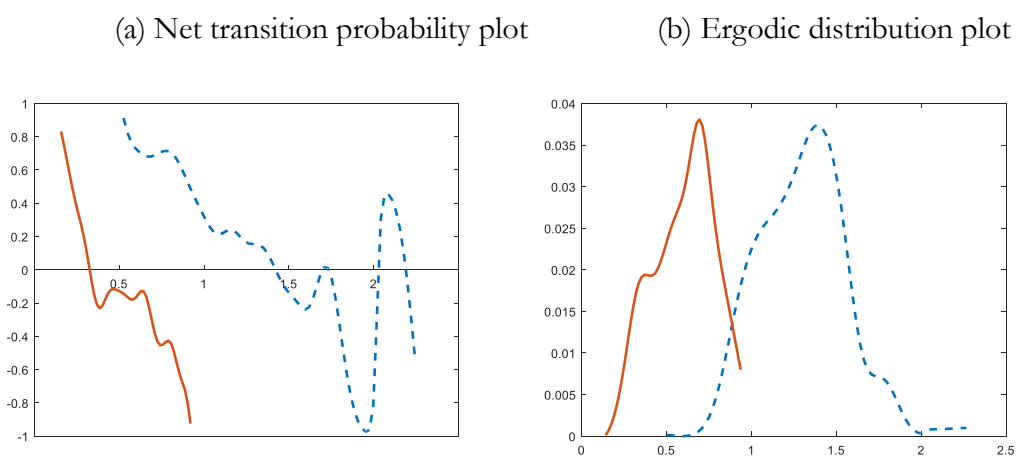

(a) Net transition probability plot    (b) Ergodic distribution plot

Figure 10 The urban and rural income distribution dynamics of the western prefectures

for the period 1999-2013

Note: The dash line and solid line show the urban and rural income, respectively

### 5.4 The income distribution dynamics of the poorest urban and rural areas

As the income growth of the poorest areas is more directly related with poverty reduction, and is also more relevant with social and political stability; thus both policymakers and researchers are more concerned about the dynamic behaviors of the poorest areas. In this subsection, we examine the income distribution dynamics of the poorest one third (measured by the income in the initial year of 1999) in the prefectural urban and rural areas.

Figure 11 plots the income distribution dynamics of the poorest one third prefectural urban areas for the period 1999-2013. The three-dimensional plot (the panel (a) in Figure 11) and contour map (the panel (b) in Figure 11) show that there is strong persistence in the lower income end, but high mobility in higher income end. In particular, the distribution of transition probability density above the RI value of 1.2 is almost parallel to the $t$ axis. This implies the existence of strong downward moving tendency in the high income end. As indicated previously, this implies convergence. The net transitional probability plot (the panel (c) in Figure 11) further confirms the convergence trend in the poorest urban areas. The prefectural urban areas with the RI values above 1.1 have strong tendency to move down in the intra-distribution dynamics, while those with the RI values below 1.1 have a net tendency to move up.

(a) Three-dimensional plot        (b) Contour map

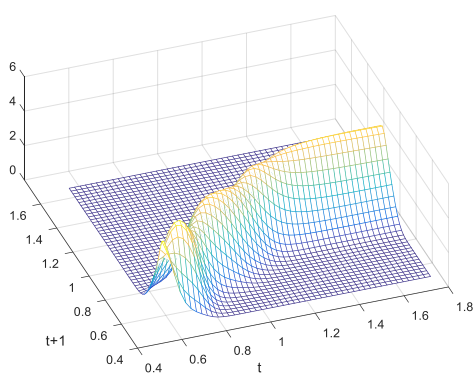 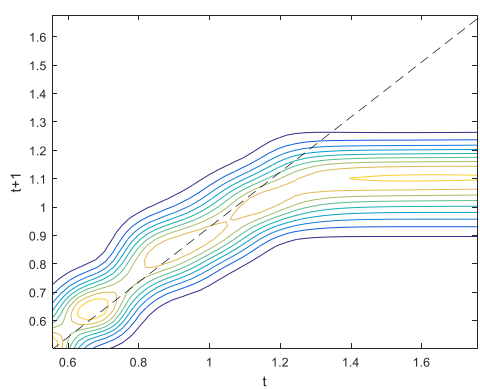

(c) Net transition probability plot        (d) Ergodic distribution plot

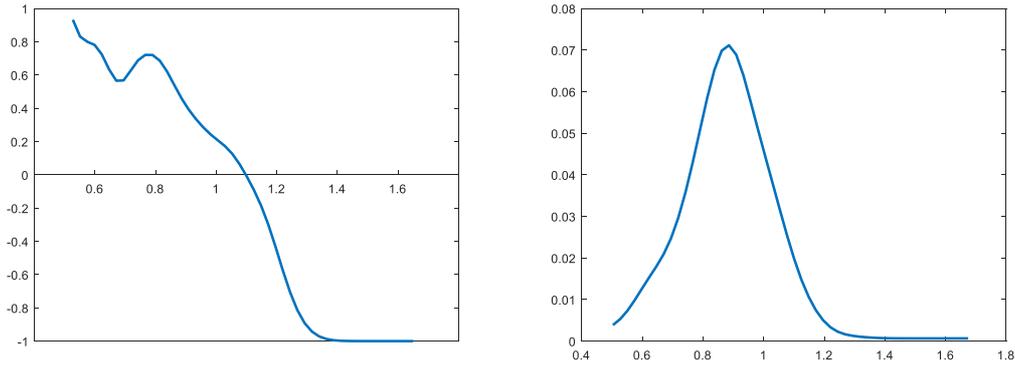

Figure 11 The income distribution dynamics of the poorest prefectural urban areas for

the period 1999-2013

The ergodic distribution in the panel (c) in Figure 11 indicates that the poorest urban areas will converge into a unique peak around 0.84 times average income. This implies the existence of poverty trap in the prefectural urban areas. The initially poor urban areas will remain poor in the future. They have little chance to escape from their current poverty traps. In the poorest one third prefectural urban areas, 45.7 and 40.4 percent of them locate in the central and western regions, only 13.9 percent locates in the eastern region. Thus the poverty traps in urban areas are a geographical phenomenon.

Figure 12 plots the income distribution dynamics of the poorest one third of prefectural rural areas for the period 1999-2013. The results are quite similar to that of the poorest urban areas. Strong downward moving tendency can be observed in the region above the RI value of 0.45 in the panel (b) of Figure 12, and relatively high persistence below the RI value of 0.45.

(a) Three-dimensional plot  (b) Contour map

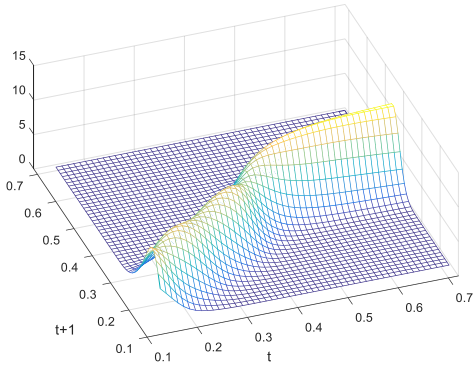 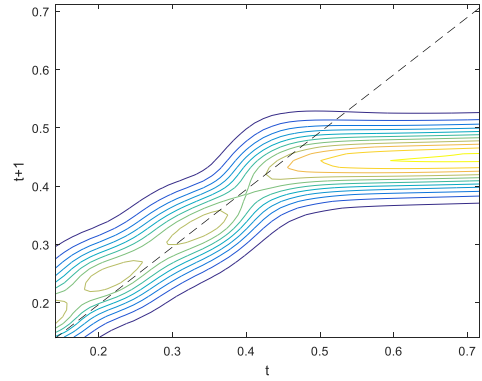

(c) Net transition probability plot      (d) Ergodic distribution plot

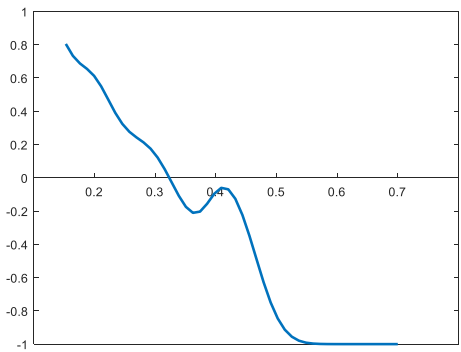 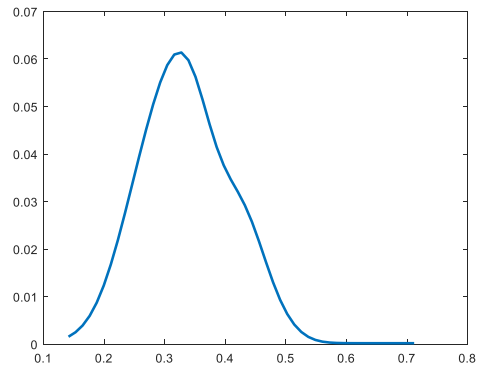

Figure 12 The income distribution dynamics of the poorest prefectural rural areas for the

period 1999-2013

    The net transition curve in the panel (c) of Figure 12 has a broadly negative slope and only one intersection with the horizontal axis. Thus strong tendency of convergence can be observed from both the contour plot and the net transitional probability plot. Keeping this transitional dynamics unchanged, as indicated by the ergodic distribution in the panel (d) of Figure 12, the poorest rural areas will cluster into a unique peak around 0.32 times average income level in the long-run. This implies that poverty trap also exists in the prefectural rural areas. In our sample, most of the poorest prefectural rural areas are in the western and central regions (65.1 and 24.8 respectively), while only 10.1 percent of them are in the eastern region. Poverty trap in rural areas is geographically

more concentrated than that in urban areas.

## 6. Conclusions and discussions

The issue of income disparity in China has attracted considerable attention and is getting worse as income gaps are getting wider. Geographically, the income inequality in China has two dimensions, namely urban-rural gap and regional divide. Different from the existing studies, this paper examines the dynamic behavior of the incomes for Chinese prefectural urban and rural areas with a distribution dynamics approach, which have advantages in analyzing persistence, polarization, and convergence clubs.

The results show that persistence and immobility are the dominant characteristics in the intradistribution dynamics of both urban and rural income in China. However, the urban and rural incomes converge to clubs differentiated by income levels. The urban incomes converge to the peak around the 1.1 times average income level, while the rural incomes converge to the peak around 0.48 times average. Examination on the three regional groups shows that urban-rural gaps (polarization) exist in all three regions, namely the Chinese eastern, central and western, but more salient in central region. Rural incomes in all three regions have strong tendency of convergence, but no simple convergence can be observed from the income distribution dynamics of the urban areas.

Significant poverty traps can be observed in both urban and rural areas. These areas converge into low income levels and have little chance to move to higher income groups. Most of the poorest prefectural urban and rural areas locate in the western and central regions. The poorest urban areas are almost equally distributed in the central and western regions, but most of the poorest rural areas cluster in the western region. Only a small number of them locate in the eastern region. Therefore, the poverty traps in China have significant geographical characteristics.

Since the Chinese government has launched many programs to reduce urban-rural disparity, and regional divide, such as the Western Development Program in 2000, the Rise of Central China Program and the Revitalization of the Northeast Program in 2004, the Socialist New Countryside Development Program in 2000, abolishment of agriculture taxes in 2005, and the goal of "Building a Harmonious Society" in 2006. However, these policies do not improve the intradistribution mobility in the poorest groups. Thus more policy interventions are required to take people out of poverty traps. As the traditional policy prescriptions, such as regional development programs and financial subsidies, are not effective in terms of reducing inequality. Less-traditional policy areas, such as promoting more migration, land property right reform and education equality, should be considered in the future.

**Acknowledgements**

Work of this paper was supported by the China National Social Science Foundation (No. 15ZDA054), Humanities and Social Sciences of Ministry of Education Planning Fund (16YJA790050), and the National Natural Science Foundation of China ((Nos. 71473105, 71273261 and 71573258).

**Reference**

Azariadis, C., and J. Stachurski, 2005. "Poverty Traps." *Handbook of Economic Growth* 1:295-384.

Bulli, S., 2001. "Distribution Dynamics and Cross-Country Convergence: A New Approach." *Scottish Journal of Political Economy* 48(2):226-243.

Cao, S., X. Wang, and G. Wang. 2009. "Lessons Learned from China's Fall into the Poverty Trap." *Journal of Policy Modeling* 31(2):298-307.

Cheng, F., X., Zhang, and S. Fan. 2002. "Emergence of Urban Poverty and Inequality in China: Evidence from Household Survey." *China Economic Review* 13(4):430-443.

CSY (various years). *The China Statistical Yearbook for Regional Economy*. Beijing: China Statistics Press.

Demurger, S., M. Fournier, and S. Li. 2006. "Urban Income Inequality in China Revisited (1988–2002)." *Economics Letters* 93(3): 354–359.

Fang, Y., and W. Zou. 2014. "Neighborhood Effects and Regional Poverty Traps in Rural China." *China & World Economy* 22(1), 83-102.

Jalan, J., and M. Ravallion. 2002. "Geographic Poverty Traps? A Micro Model of Consumption Growth in Rural China." *Journal of Applied Econometrics* 17(4): 329-346.

Johnson, P. A., 2005. "A Continuous State Space Approach to 'Convergence by Parts'." *Economics Letters* 86(3):317-321.

Kanbur, R. and X. Zhang. 2005. "Fifty Years of Regional Inequality in China: A Journey through Central Planning, Reform, and Openness." *Review of Development Economics* 9(1): 87–106.

Knight, J., Shi, L., and D. Quheng. 2009. "Education and the poverty trap in rural China:

Setting the trap." *Oxford Development Studies* 37(4):311-332.

Knight, J., L. Shi, and D. Quheng, 2010. "Education and the Poverty Trap in Rural China: Closing the Trap." *Oxford Development Studies*, 38(1):1-24.

Kwong, T. 1994. "Markets and urban-rural inequality in China." *Social Science Quarterly* 75(4): 821–837.

Lardy, N. 1978. "*Economic Growth and Distribution in China.*" New York: Cambridge University Press.

Li, S., C. Luo, and T. Sicular. 2013. "Overview: Income Inequality and Poverty in China, 2002–2007." In *Rising Inequality in China: Challenge to the Harmonious Society*, eds. Li Shi, Hiroshi Sato, and Terry Sicular, 24-59. Cambridge and New York: Cambridge University Press.

Lyons, T. 1991. "Interprovincial Disparities in China: Output and Consumption, 1952–1987." *Economic Development and Cultural Change* 39(3): 471–506.

Nolan, P. 1979. "Inequality of Income between Town and Countryside in the People's Republic of China in the Mid-1950s." *World Development* 7(4–5): 447–465.

Quah, D. T. 1993. "Empirical Cross-Section Dynamics in Economic Growth." *European Economic Review* 37(2):426-434.

Quah, D. T. 1996a. "Convergence Empirics across Economies with (Some) Capital Mobility." *Journal of Economic Growth* 1(1):95-124.

Quah, D. T. 1996b. "Twin Peaks: Growth and Convergence in Models of Distribution Dynamics." *The Economic Journal* 106(437):1045-1055.

Quah, D. T. 1997. "Empirics for Growth and Distribution: Stratification, Polarization,

and Convergence Clubs." *Journal of Economic Growth* 2(1):27-59.

Ravallion, M. and S. Chen. 2007. "China's (uneven) progress against poverty." *Journal of Development Economics* 82(1): 1–42.

Sicular, T., Y. Ximing, B. Gustafsson, and L. Shi. 2007. "The Urban–Rural Income Gap and Inequality in China." *Review of Income and Wealth 53*(1):93-126.

Silverman, B. W. 1986. "*Density Estimation for Statistics and Data Analysis.*" CRC press.

Tsui, K. 1991. "China's Regional Inequality, 1952–1985." *Journal of Comparative Economics* 15(1): 1–21.

Wan, G. 2007. "Understanding Regional Poverty and Inequality Trends in China: Methodological Issues and Empirical Findings." *Review of Income and Wealth*, *53*(1):25-34.

Wan, G. and Z. Zhou. 2005. "Income Inequality in Rural China: Regression-Based Decomposition Using Household Data." *Review of Development Economics* 9(1): 107–120.

Wan, G., and I. Sebastian. 2011. "Poverty in Asia and the Pacific: An update." *Asian Development Bank Economics Working Paper*, (267).

Wang, Y., Y. Liu, Y. Li, and T. Li. 2016. "The Spatio-Temporal Patterns of Urban–Rural Development Transformation in China since 1990." *Habitat International* 53:178-187.

Wang, C., G. Wan, and D. Yang. 2014. "Income Inequality in the People's Republic of China: Trends, Determinants, and Proposed Remedies." *Journal of Economic Surveys 28*(4):686-708.

Yang, D. and H. Zhou. 1999. "Rural-Urban Disparity and Sectorial Labour Allocation in China." *Journal of Development Studies* 35(3): 105–133.

Zhang, Q. and H. Zou. 2012. "Regional inequality in contemporary China." *Annals of*

*Economics and Finance* 13(1): 113–137.

Zhao, R. 1993. "Three features of the distribution of income during the transition to reform." In K. Griffin and R. Zhao (eds), *The Distribution of Income in China*. New York: St. Martin's Press.